\documentclass[aps,prl,twocolumn,superscriptaddress,showpacs,preprintnumbers,amsmath,amssymb]{revtex4}
\usepackage{graphicx} % Include figure files
\usepackage{dcolumn}  % Align table columns on decimal point

\begin{document}

%\vspace*{-3\baselineskip}
%\resizebox{!}{3cm}{\includegraphics{belle.eps}}

%\preprint{\vbox{ \hbox{   }
%                 \hbox{BELLE-CONF-0205}
%                 \hbox{ICHEP02 Parallel 7 or 8}
%                 \hbox{ABS692}
%                 \hbox{hep-ex nnnn, if available}
%}}

\title{ \quad\\[0.5cm]  Measurements of the $D_{sJ}$ resonance
properties}

%\input{author}

%%% Paper:    e+e- -> D_sJ
%%% Journal:  Physical Review Letters
%%% Contacts: Y. Mikami (mikami@bmail.kek.jp)
%%% Non-responding authors or those who said NO are commented out.
%%% ====================================================================
%%% Click the RELOAD button on your web browser to see the updated file.
%%% ====================================================================
%%% Use \input{author} to insert this material into your latex file.
%%%%% Force institutions to appear in alphabetical order when typeset.
\affiliation{Aomori University, Aomori}
\affiliation{Budker Institute of Nuclear Physics, Novosibirsk}
\affiliation{Chiba University, Chiba}
%%%\affiliation{Chuo University, Tokyo}
\affiliation{University of Cincinnati, Cincinnati, Ohio 45221}
%%%\affiliation{University of Frankfurt, Frankfurt}
\affiliation{Gyeongsang National University, Chinju}
\affiliation{University of Hawaii, Honolulu, Hawaii 96822}
\affiliation{High Energy Accelerator Research Organization (KEK), Tsukuba}
\affiliation{Hiroshima Institute of Technology, Hiroshima}
\affiliation{Institute of High Energy Physics, Chinese Academy of Sciences, Beijing}
\affiliation{Institute of High Energy Physics, Vienna}
\affiliation{Institute for Theoretical and Experimental Physics, Moscow}
\affiliation{J. Stefan Institute, Ljubljana}
\affiliation{Kanagawa University, Yokohama}
\affiliation{Korea University, Seoul}
%%%\affiliation{Kyoto University, Kyoto}
\affiliation{Kyungpook National University, Taegu}
\affiliation{Institut de Physique des Hautes \'Energies, Universit\'e de Lausanne, Lausanne}
\affiliation{University of Ljubljana, Ljubljana}
\affiliation{University of Maribor, Maribor}
\affiliation{University of Melbourne, Victoria}
\affiliation{Nagoya University, Nagoya}
\affiliation{Nara Women's University, Nara}
%%%\affiliation{National Kaohsiung Normal University, Kaohsiung}
%%%\affiliation{National Lien-Ho Institute of Technology, Miao Li}
\affiliation{Department of Physics, National Taiwan University, Taipei}
\affiliation{H. Niewodniczanski Institute of Nuclear Physics, Krakow}
\affiliation{Nihon Dental College, Niigata}
\affiliation{Niigata University, Niigata}
\affiliation{Osaka City University, Osaka}
\affiliation{Osaka University, Osaka}
\affiliation{Panjab University, Chandigarh}
\affiliation{Peking University, Beijing}
\affiliation{Princeton University, Princeton, New Jersey 08545}
%%%\affiliation{RIKEN BNL Research Center, Upton, New York 11973}
%%%\affiliation{Saga University, Saga}
\affiliation{University of Science and Technology of China, Hefei}
\affiliation{Seoul National University, Seoul}
\affiliation{Sungkyunkwan University, Suwon}
\affiliation{University of Sydney, Sydney NSW}
\affiliation{Tata Institute of Fundamental Research, Bombay}
\affiliation{Toho University, Funabashi}
\affiliation{Tohoku Gakuin University, Tagajo}
\affiliation{Tohoku University, Sendai}
\affiliation{Department of Physics, University of Tokyo, Tokyo}
\affiliation{Tokyo Institute of Technology, Tokyo}
\affiliation{Tokyo Metropolitan University, Tokyo}
\affiliation{Tokyo University of Agriculture and Technology, Tokyo}
\affiliation{Toyama National College of Maritime Technology, Toyama}
\affiliation{University of Tsukuba, Tsukuba}
%%%\affiliation{Utkal University, Bhubaneswer}
\affiliation{Virginia Polytechnic Institute and State University, Blacksburg, Virginia 24061}
\affiliation{Yokkaichi University, Yokkaichi}
\affiliation{Yonsei University, Seoul}
  \author{Y.~Mikami}\affiliation{Tohoku University, Sendai} % Tohoku
  \author{K.~Abe}\affiliation{High Energy Accelerator Research Organization (KEK), Tsukuba} % KEK
% \author{K.~Abe}\affiliation{Tohoku Gakuin University, Tagajo} % TohokuGakuin
% \author{N.~Abe}\affiliation{Tokyo Institute of Technology, Tokyo} % TIT
  \author{T.~Abe}\affiliation{High Energy Accelerator Research Organization (KEK), Tsukuba} % KEK
% \author{I.~Adachi}\affiliation{High Energy Accelerator Research Organization (KEK), Tsukuba} % KEK
  \author{H.~Aihara}\affiliation{Department of Physics, University of Tokyo, Tokyo} % Tokyo
% \author{K.~Akai}\affiliation{High Energy Accelerator Research Organization (KEK), Tsukuba} % KEK
% \author{M.~Akatsu}\affiliation{Nagoya University, Nagoya} % Nagoya
  \author{M.~Akemoto}\affiliation{High Energy Accelerator Research Organization (KEK), Tsukuba} % KEK
% \author{M.~Asai}\affiliation{Hiroshima Institute of Technology, Hiroshima} % Hiroshima
  \author{Y.~Asano}\affiliation{University of Tsukuba, Tsukuba} % Tsukuba
  \author{T.~Aso}\affiliation{Toyama National College of Maritime Technology, Toyama} % Toyama
% \author{V.~Aulchenko}\affiliation{Budker Institute of Nuclear Physics, Novosibirsk} % BINP
  \author{T.~Aushev}\affiliation{Institute for Theoretical and Experimental Physics, Moscow} % ITEP
  \author{S.~Bahinipati}\affiliation{University of Cincinnati, Cincinnati, Ohio 45221} % Cincinnati
  \author{A.~M.~Bakich}\affiliation{University of Sydney, Sydney NSW} % Sydney
% ...
% Ljubljana
% \author{D.~Z\"urcher}\affiliation{Institut de Physique des Hautes \'Energies, Universit\'e de Lausanne, Lausanne} % Lausanne
\collaboration{The Belle Collaboration}

\noaffiliation

\begin{abstract}
We report measurements of the properties of the
$D_{sJ}^+(2317)$ and $D_{sJ}^+(2457)$ resonances 
produced in continuum $e^+ e^-$ annihilation
near $\sqrt{s}=10.6\,\mathrm{GeV}$.  The analysis
is based on an $86.9\,\mathrm{fb^{-1}}$ data sample collected
%at and 60~MeV below the $\Upsilon(4S)$  resonance with
with the  Belle detector at KEKB.  
We determine the masses to be 
$M(D_{sJ}^+(2317)) = 2317.2 \pm 0.5(\mathrm{stat}) 
\pm 0.9(\mathrm{syst})\,\mathrm{MeV}/c^2$ 
and 
$M(D_{sJ}^+(2457))=2456.5 \pm 1.3(\mathrm{stat}) 
\pm 1.3(\mathrm{syst})\,\mathrm{MeV}/c^2$. 
We observe the radiative decay mode  $D_{sJ}^+(2457) \to
 D_s^+ \gamma$ and the dipion decay mode  $D_{sJ}^+(2457) \to
 D_s^+ \pi^+ \pi^-$, and determine their branching fractions.
No corresponding decays are observed for the $D_{sJ}(2317)$ state. 
These results are consistent with the spin-parity assignments of $0^+$ for 
the $D_{sJ}(2317)$ and  $1^+$ for the $D_{sJ}(2457)$.

\end{abstract}


\maketitle
\tighten

{\renewcommand{\thefootnote}{\fnsymbol{footnote}}}
\setcounter{footnote}{0}

The narrow $D_s \pi^0$ resonance at 2317 MeV/$c^2$, recently observed 
by the BaBar collaboration~\cite{babar_dspi0}, is naturally interpreted 
as a P-wave excitation of the $c\bar{s}$ system. 
The observation of a nearby  and narrow $D_s^{*} \pi^0$ resonance 
by the  CLEO collaboration~\cite{cleo_dspi0} supports this view, 
since the mass difference of the two observed states is consistent with 
the expected hyperfine splitting for a P-wave doublet with total 
light-quark angular momentum $j=1/2$~\cite{bardeen,pwave-doublets}.
The observed masses are, however, considerably lower than potential model 
predictions~\cite{bartelt} and similar to those of the 
$c\bar{u}$ $j=1/2$ doublet states recently reported by Belle~\cite{kuzmin}.
This has led to speculation that the new $D_s^{(*)} \pi^0$ resonances, 
which we denote $D_{sJ}$, may be exotic 
mesons~\cite{cahn,lipkin,beveren,hou,fazio,godfrey}.
Measurements of the $D_{sJ}$ quantum numbers and branching 
fractions (particularly those for radiative decays),
will play an important role in determining the nature of these states.

In this paper we report measurements of the $D_{sJ}$ masses, 
widths and branching fractions 
using a sample of $e^+ e^- \to c\bar{c}$
events collected with the Belle detector~\cite{Belle} at 
the KEKB collider~\cite{KEKB}.

We reconstruct $D_s^+$ mesons using the decay chain 
$D_s^+ \rightarrow \phi \pi^+$ and $\phi \rightarrow K^+ K^-$. 
To identify kaons or pions, 
we form a likelihood for each track, $\mathcal{L}_{K(\pi)}$,
from  $dE/dx$ measurements in a 50-layer central drift chamber,
the responses from aerogel threshold \v{C}erenkov counters, 
and time-of-flight scintillation counters. 
The kaon likelihood ratio, 
$P(K/\pi) = \mathcal{L}_K /(\mathcal{L}_K + \mathcal{L}_\pi)$, has values 
between 0 (likely to be a pion) and 1 (likely to be a kaon).

For $\phi\rightarrow K^+K^-$ candidates we use oppositely-charged track pairs 
where one track has $P(K/\pi)> 0.5$ and the other has  $P(K/\pi)> 0.2$, 
and with
a $K^+ K^-$ invariant mass that is within 
$10\,\mathrm{MeV}/c^2\,(\sim 2.5\sigma)$ 
of the nominal $\phi$ mass. 
We define the $\phi$ helicity angle $\theta_H$ to be the angle between the 
direction of the $K^+$ and the $D_s^+$ in the  $\phi$ rest frame.  For signal
events this has a $\cos^2{\theta_H}$ distribution,  while for background
it is flat; we require $|\cos\theta_H|> 0.35$.

We reconstruct $D_s^+$ candidates by combining a $\phi$ candidate with
a $\pi^+$, which is a charged track with $P(K/\pi) < 0.9$, and
requiring $M(\phi \pi^+)$ to be within 
$10\,\mathrm{MeV}/c^2\,(\sim 2\sigma)$ of
the nominal $D_s^+$  mass.    We use the $D_s^+$ sideband regions 
$1920 < M(\phi \pi^+) < 1940\,\mathrm{MeV}/c^2$ and 
$1998 < M(\phi \pi^+) < 2018\,\mathrm{MeV}/c^2$
for background studies. 

For $\pi^0$ reconstruction, we use photons with the $e^+ e^-$ rest frame (CM) 
energies greater than $100\,\mathrm{MeV}$ and select $\gamma \gamma$ pairs 
that have an invariant mass  
$M(\gamma\gamma)$ within $10\,\mathrm{MeV}/c^2\,(\sim 2\sigma)$ of 
the $\pi^0$ mass. 
For background studies we use the $\pi^0$ sideband regions
$105 \leq M_{\gamma\gamma} \leq 115\,\mathrm{MeV}/c^2$ 
and $155 \leq M_{\gamma\gamma} \leq 165\,\mathrm{MeV}/c^2$. 

We reconstruct $D_s^{*+}$ in the $D_s^+ \gamma$ final state.
We use photons with CM energies greater than 
$100\,\mathrm{MeV}$ and require
$D_s^{*+}$ candidates to satisfy  
$127 \leq \Delta M(D_s^+ \gamma)  \leq 157\,\mathrm{MeV}/c^2\,(\sim 3\sigma)$,
where $\Delta M(D_s^+\gamma) = M(D_s^+\gamma) - M_{D_s^+}$.  
The $D_s^{*+}$ sideband regions are defined as 
$87 \leq \Delta M(D_s^+ \gamma) \leq 117\,\mathrm{MeV}/c^2$ 
and $167 \leq \Delta M(D_s^+ \gamma) \leq 197\,\mathrm{MeV}/c^2$.

\begin{figure}[t]
\begin{minipage}{4.25cm}
\begin{center}
\vspace*{-5mm}
\includegraphics[width=4.7cm,clip]{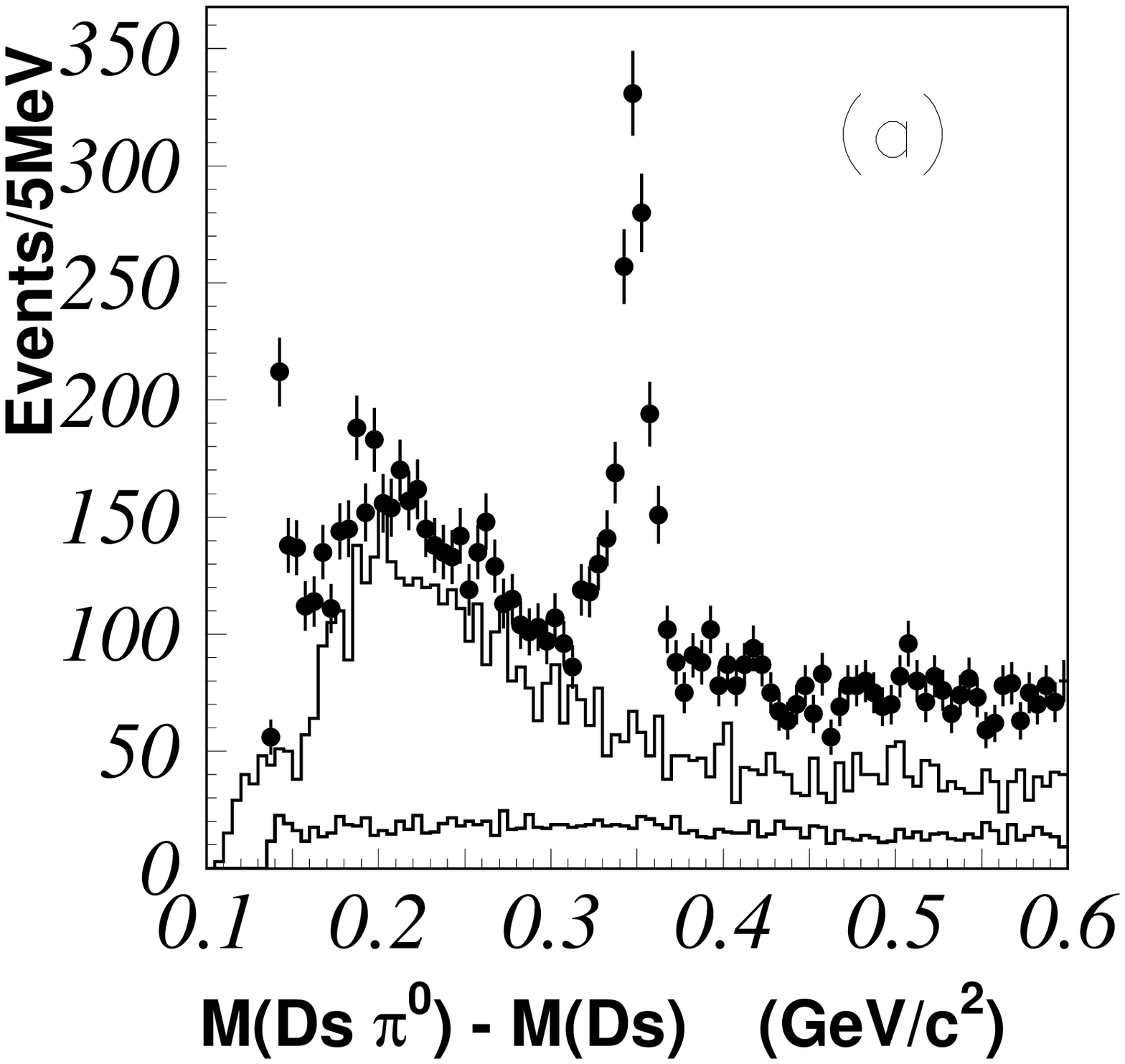}
\end{center}
\end{minipage}
\begin{minipage}{4.25cm}
\begin{center}
\vspace*{-5mm}
\includegraphics[width=4.7cm,clip]{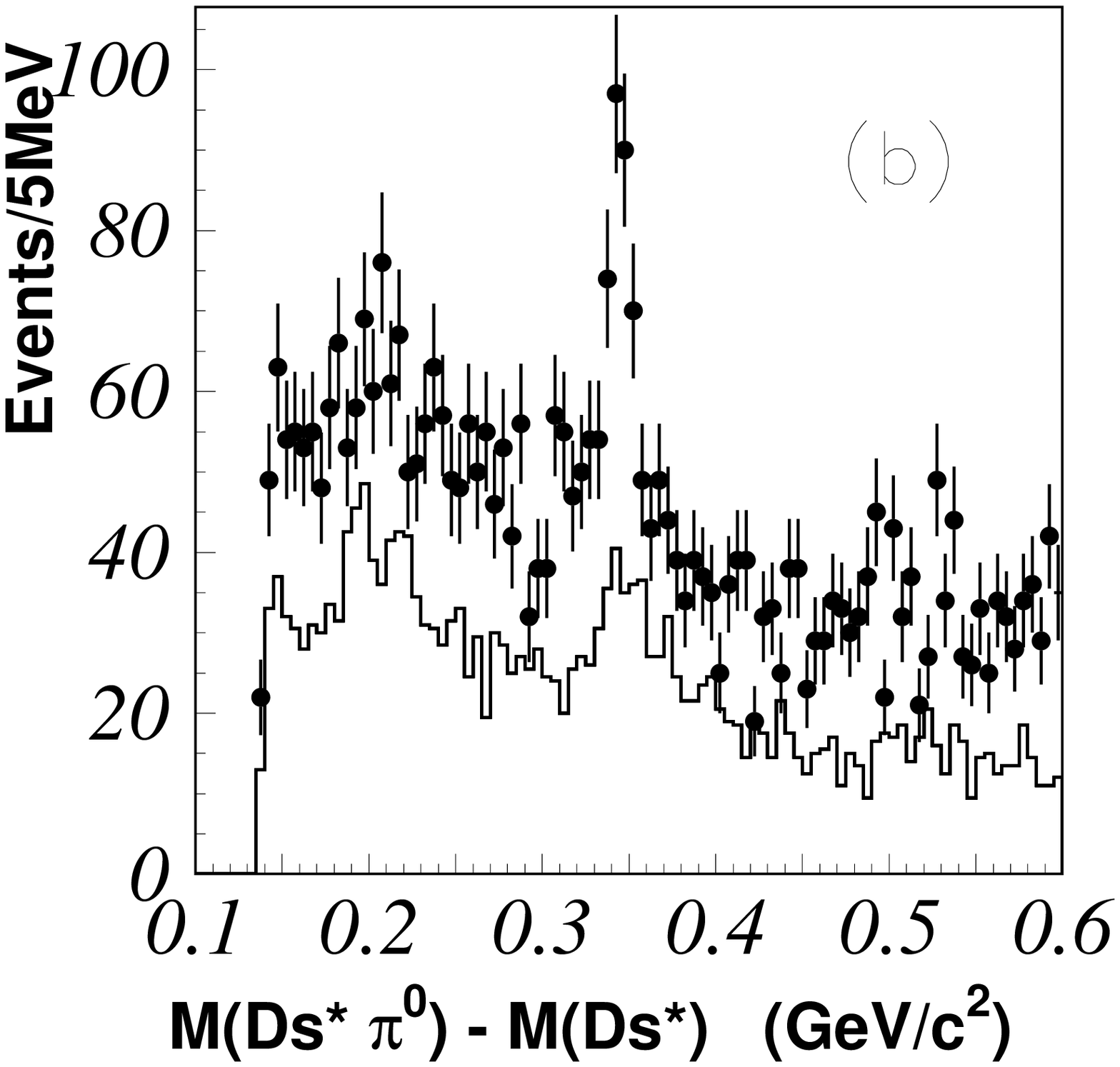}
\end{center}
\end{minipage}
 \caption{(a) The $\Delta M(D_s^+ \pi^0)$  distribution. 
        Data from the $D_s^+$ (solid) and $\pi^0$
 (dashed) sideband regions are also shown.
(b) The $\Delta M(D_s^{*+} \pi^0)$ distribution.  
Data  from the $D_s^*$ sideband (histogram) region is also shown. }
\label{peaks}
\end{figure}

The $\Delta M(D_s^+ \pi^0) = M(D_s^+ \pi^0)-M_{D_s^+}$
mass-difference distribution 
for $D_s^+ \pi^0$ combinations with $p^*(D_s^+ \pi^0) > 3.5\,\mathrm{GeV}/c$
is shown in Fig.~\ref{peaks}(a).  Here, and in analyses of other $D_{sJ}$
states and modes, we require the CM momentum 
to satisfy $p^*(D_{sJ}) > 3.5\,\mathrm{GeV}/c$ 
to remove contributions from $B \bar B$ events. 
We do not remove multiple candidates in the subsequent analysis.
Also shown are the distributions for the
$D_s^+$ (solid) and $\pi^0$  (dashed) sideband regions.
The prominent peak in the figure corresponds to the
$D_{sJ}(2317)\to D_s^+\pi^0$ signal; the peak at small $\Delta M$ values
is due to $D_s^{*+}(2112) \to D_s^+ \pi^0$. 
No peak is seen in the sideband distributions.  

Figure~\ref{peaks}(b) shows 
the $\Delta M (D_s^{*+} \pi^0) = M(D_s^{*+} \pi^0)-M_{D_s^{*+}}$ 
distribution for $p^*(D_s^{*+} \pi^0) > 3.5\,\mathrm{GeV}/c$,
where a peak corresponding to $D_{sJ}(2457)\to D_s^{*+} \pi^0$
is evident. Also shown is the distribution for the $D_s^{*+}$ sideband
region, where we notice the presence of a wider peak in the $D_{sJ}(2457)$
region. The $\Delta M (D_s^{*+} \pi^0)$  distributions for the $D_s^+$ 
and $\pi^0$ sideband regions show no such peak.

To study the expected signal shape and detection efficiencies,
and determine the level of cross-feed between the two states, 
we use a Monte Carlo simulation that treats the $D_{sJ}(2317)$
as a scalar particle with mass $2317\,\mathrm{MeV}/c^2$ decaying to 
$D_s^+ \pi^0$ 
and the $D_{sJ}(2457)$ as an axial-vector particle with mass 
$2457\,\mathrm{MeV}/c^2$ decaying 
to $D_s^{*+} \pi^0$. Zero intrinsic width is assigned to both states. 
We find that the $D_{sJ}(2317)$ produces a peak of 
width $7.1 \pm 0.2\,\mathrm{MeV}/c^2$
in the $\Delta M(D_s^+\pi^0)$ distribution at its nominal mass, and a 
broader reflection peak (of width $12.3 \pm 1.8\,\mathrm{MeV}/c^2$) 
at a mass of $8\,\mathrm{MeV}/c^2$ above the $D_{sJ}(2457)$ peak. 
This latter peak
corresponds to a $D_s^+$ and $\pi^0$ from a $D_{sJ}(2317)$ decay that
are combined with a random photon that passes the 
$|M(D_s^+ \gamma)-M_{D_s^{*+}}| < 15\,\mathrm{MeV}/c^2$ requirement.
(We refer to this as ``feed-up background.'')
The $D_{sJ}(2457)$ produces a peak of width $6.0\pm 0.2\,\mathrm{MeV}/c^2$ 
at its nominal mass
and a broader peak (of width $19.5 \pm 3.6\,\mathrm{MeV}/c^2$), 
also at its nominal 
mass.  The latter peak is due 
to events in which the photon from $D_s^{*+} \to D_s^+ \gamma$ is missed, 
and a random photon is reconstructed in its place 
(referred to as the ``broken-signal background'').
In addition, the $D_{sJ}(2457)$ produces a reflection in the $D_s^+
\pi^0$ mass distribution with width $14.9 \pm 0.8\,\mathrm{MeV}/c^2$ 
at a mass of $4\,\mathrm{MeV}/c^2$ below the  
$D_{sJ}(2317)$ peak (referred to as ``feed-down background'').

While we must depend on the MC for separating the signal peak and the
feed-down background in the $D_{sJ}(2317)$ region,
the feed-up and broken-signal backgrounds for the $D_{sJ}(2457)$ region 
occur when $D_s^{*+} \pi^0$ combinations are formed from candidates 
in the $D_s^{*+}$ mass sideband. This is evident in
Fig.~\ref{peaks}(b). 

\begin{figure}[h]
\begin{minipage}{4.25cm}
\begin{center}
\vspace*{-5mm}
\includegraphics[width=4.7cm,clip]{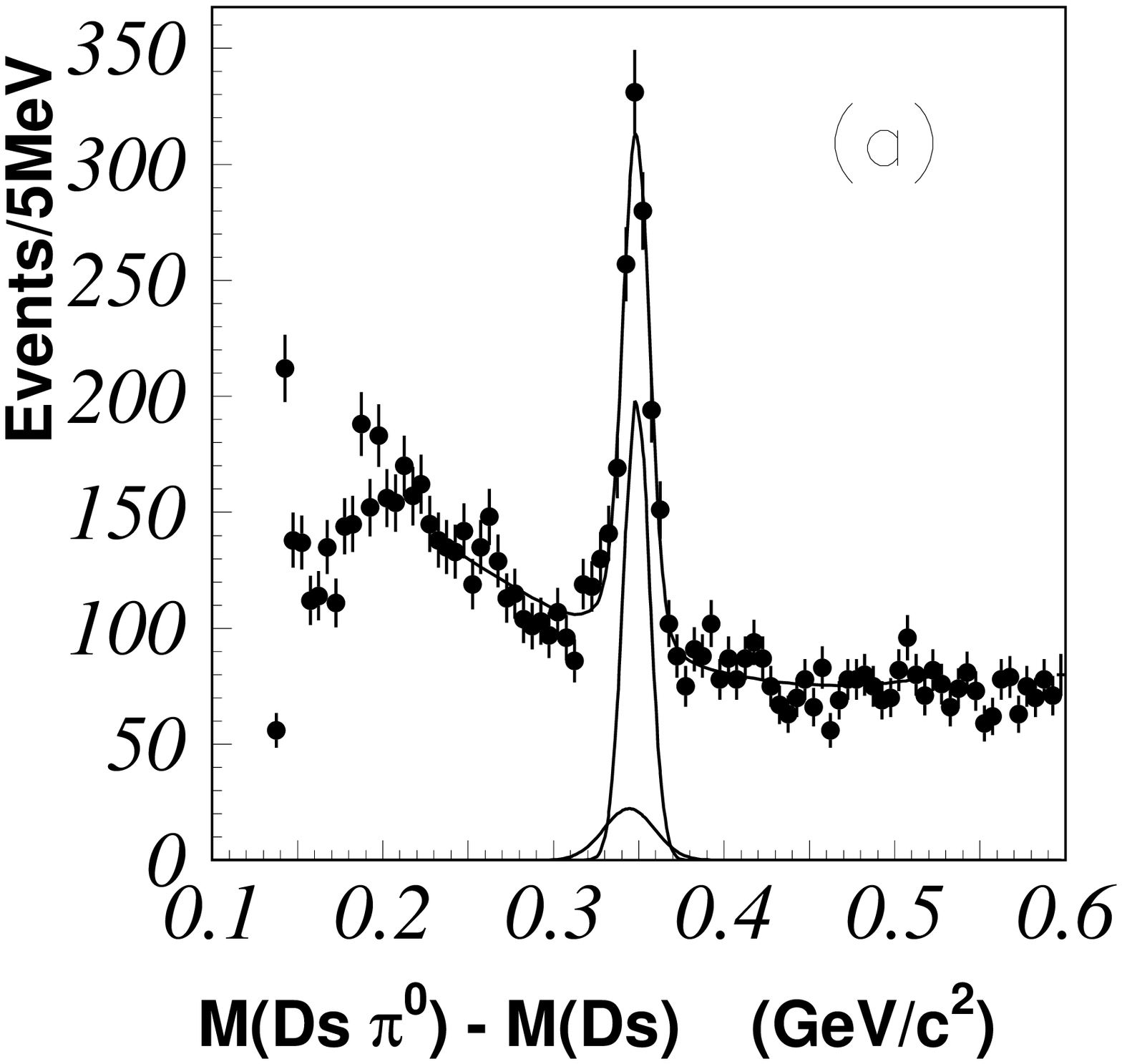}
\end{center}
\end{minipage}
\begin{minipage}{4.25cm}
\begin{center}
\vspace*{-5mm}
\includegraphics[width=4.7cm,clip]{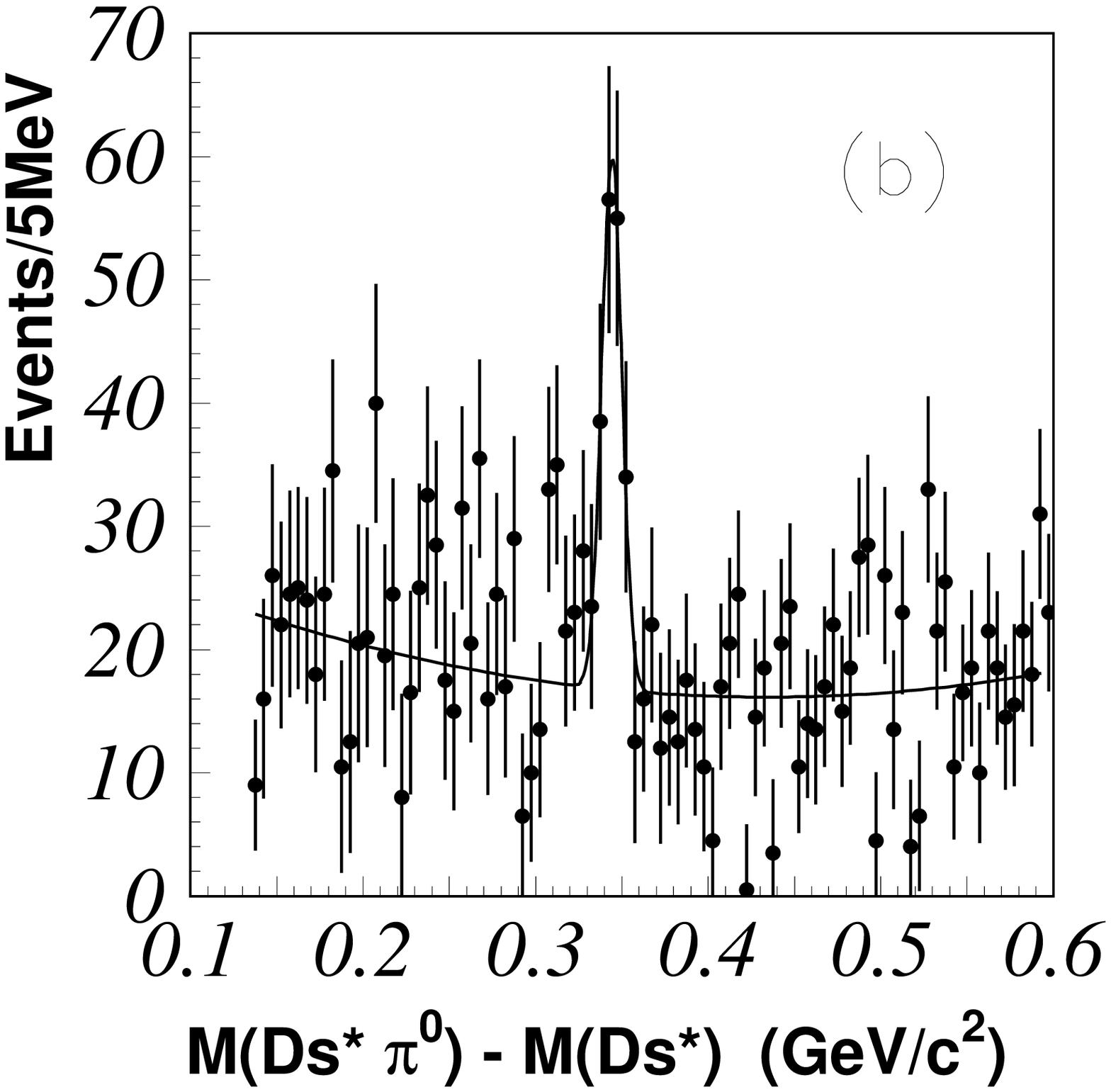}
\end{center}
\end{minipage}
 \caption{(a) The $\Delta M(D_s^+ \pi^0)$ distribution. 
The narrow Gaussian peak is the fitted $D_{sJ}(2317)$ signal, whereas  the 
wider Gaussian peak is the feed-down background.  
 (b) The $\Delta M(D_s^{*+} \pi^0)$  distribution  
   after bin-by-bin subtraction of the $D_s^{*+}$ 
  sideband from  the $D_s^{*+}$ signal distribution. 
   The curve is the fit result. }
\label{masses}
\end{figure}

Figure~\ref{masses}(b) shows the sideband-subtracted 
$\Delta M(D_s^{*+} \pi^0)$ distribution together with the results
of a fit that uses a Gaussian to represent the $D_{sJ}(2457)$ signal
and a second-order polynomial for the background. 
The fit gives a signal yield of $126 \pm 25$ events with a peak 
value of $\Delta M = 344.1 \pm 1.3\,\mathrm{MeV}/c^2$ (corresponding to
$M = 2456.5 \pm 1.3\,\mathrm{MeV}/c^2$).  The width from the fit, 
$\sigma = 5.8 \pm 1.3~\mathrm{MeV}/c^2$, is consistent with  MC
expectations for a zero intrinsic width particle.

Figure~\ref{masses}(a) shows the fit result for the
$D_{sJ}(2317)$.  Here both the signal and the feed-down background 
are represented as Gaussian shapes  modeled from the MC.  
The mean and $\sigma$ of the feed-down component are fixed according to 
the MC and normalized by the measured $D_{sJ}(2457)$ yield.  
A third-order polynomial
is used to represent the non-feed-down background.  The fit gives a   
yield of $761 \pm 44$ events and a peak $\Delta M$ value of 
 $348.7 \pm 0.5\,\mathrm{MeV}/c^2$ (corresponding to
$M = 2317.2 \pm 0.5\,\mathrm{MeV}/c^2$).  Here again, the width from
the fit, $\sigma = 7.6 \pm 0.5\,\mathrm{MeV}/c^2$, is consistent
with MC expectations for a zero intrinsic width particle. 

There are systematic errors in the measurements due to uncertainties in the:
i) $\pi^0$ energy calibration; 
ii) parameterization of the cross-feed backgrounds; 
iii) parameterization for the non-cross-feed backgrounds; 
iv) possible discrepancies 
between the input and output seen in the MC simulations; and 
v) the uncertainty
in the world average value for $M_{D_s^+}$ and $M_{D_s^{*+}}$. 

The $\pi^0$ energy calibration is studied using  
$D_s^{*+}(2112) \to D_s^+ \pi^0$ events in the same data sample. 
We measure $\Delta M = 144.3 \pm 0.1\,\mathrm{MeV}/c^2$ and 
$\sigma = 1.0 \pm 0.1\,\mathrm{MeV}/c^2$, 
which agrees well with the PDG value of 
$\Delta M = 143.8 \pm 0.4\,\mathrm{MeV}/c^2$.  
The MC, which uses the PDG value as an input, gives 
$\Delta M = 143.9 \pm 0.1\,\mathrm{MeV}/c^2$ and 
$\sigma = 1.0 \pm 0.1\,\mathrm{MeV}/c^2$. (The errors quoted here are 
statistical only). 
We attribute the difference to the $\pi^0$ energy calibration
uncertainty, and conservatively assign a $\pm 0.6\,\mathrm{MeV}/c^2$ error
to this effect. This error only contributes to the mass measurements.    
 
For the cross-feed background to the $D_{sJ}(2317)$ signal, we vary the
feed-down background parameters and the $D_{sJ}(2457)$ yield by 
$\pm 1\sigma$ and assign the variation in output values as errors. 
For the $D_{sJ}(2457)$, we determine the uncertainty
of the feed-up fraction from the difference between 
the $D_s^*$ signal region and the sideband region using the MC.
For the non-cross-feed background, we repeat the fit using a second-order
polynomial for the $D_{sJ}(2317)$ and 
a linear function for the $D_{sJ}(2457)$ and
assign the difference as errors.
Differences between the MC input and output values for the $D_{sJ}$ 
parameters can reflect possible errors arising from the choice of signal
shape and other factors in the analysis.  We assign these differences as 
errors.
   
The final results for the masses are  
\begin{eqnarray}
M(D_{sJ}(2317)) &=& 2317.2 \pm 0.5(\mathrm{stat}) \pm
0.9(\mathrm{syst})\,\mathrm{MeV}/c^2 \nonumber \\
M(D_{sJ}(2457)) &=& 2456.5 \pm 1.3(\mathrm{stat}) \pm 1.3
(\mathrm{syst})\,\mathrm{MeV}/c^2 \nonumber.
\end{eqnarray}
The $M(D_{sJ}(2317))$ result is consistent with BaBar~\cite{babar_dspi0}
and CLEO results~\cite{cleo_dspi0}.
Our $M(D_{sJ}(2457))$ value is consistent with BaBar~\cite{palano} but 
significantly lower than that from CLEO~\cite{cleo_dspi0}.
We set upper limits for the natural widths of 
$\Gamma({D_{sJ}(2317)}) \le 4.6\,\mathrm{MeV}/c^2$ and 
$\Gamma({D_{sJ}(2457)}) \le 5.5\,\mathrm{MeV}/c^2$~(90\%~C.L.), respectively.

Using the observed signal yields of
$761 \pm 44(\mathrm{stat}) \pm 30(\mathrm{syst})$ and 
$126 \pm 25(\mathrm{stat}) \pm 12(\mathrm{syst})$for the $D_{sJ}(2317)$ and 
$D_{sJ}(2457)$, and the detection efficiencies of 8.2\% and  4.7\% for 
the $D_{sJ}(2317)$ and $D_{sJ}(2457)$, we determine the ratio
\begin{eqnarray}
&&\frac{\sigma(D_{sJ}(2457) ) \cdot \mathcal{B}(D_{sJ}^+(2457) 
\to D_s^{*+} \pi^0) }
     {\sigma(D_{sJ}(2317) ) \cdot \mathcal{B}(D_{sJ}^+(2317) \to D_s^+ \pi^0)}
  \nonumber \\ 
 &=& 0.29 \pm 0.06(\mathrm{stat}) \pm 0.03(\mathrm{syst}) \nonumber. 
\end{eqnarray}
The detection efficiencies are determined from the MC assuming the same 
fragmentation function for the two states.   
The dominant source of systematic error is the systematic uncertainty
in the $D_{sJ}(2457)$ yield.

In the $D_{sJ}(2457)$ region of the $\Delta M(D_s^+ \pi^0)$ distribution,
we find $22 \pm 22$ events from a fit to a possible $D_{sJ}(2457)$ signal.
From this, we obtain the upper limit 
$$
\frac{\mathcal{B}(D_{sJ}^+(2457) \rightarrow D_s^+ \pi^0)}
     {\mathcal{B}(D_{sJ}^+(2457) \rightarrow D_s^{*+} \pi^0)} 
 \leq 0.21~(90\%~C.L.). 
$$
The decay to a pseudo-scalar pair is allowed for a 
state with a parity of $(-1)^J$. Thus, absence
of such a decay disfavors $D_{sJ}(2457)$
having $J^P$ of $0^+$ or $1^-$.

\begin{figure}[t]
\begin{minipage}{4.25cm}
\begin{center}
\vspace*{-5mm}
\includegraphics[width=4.7cm,clip]{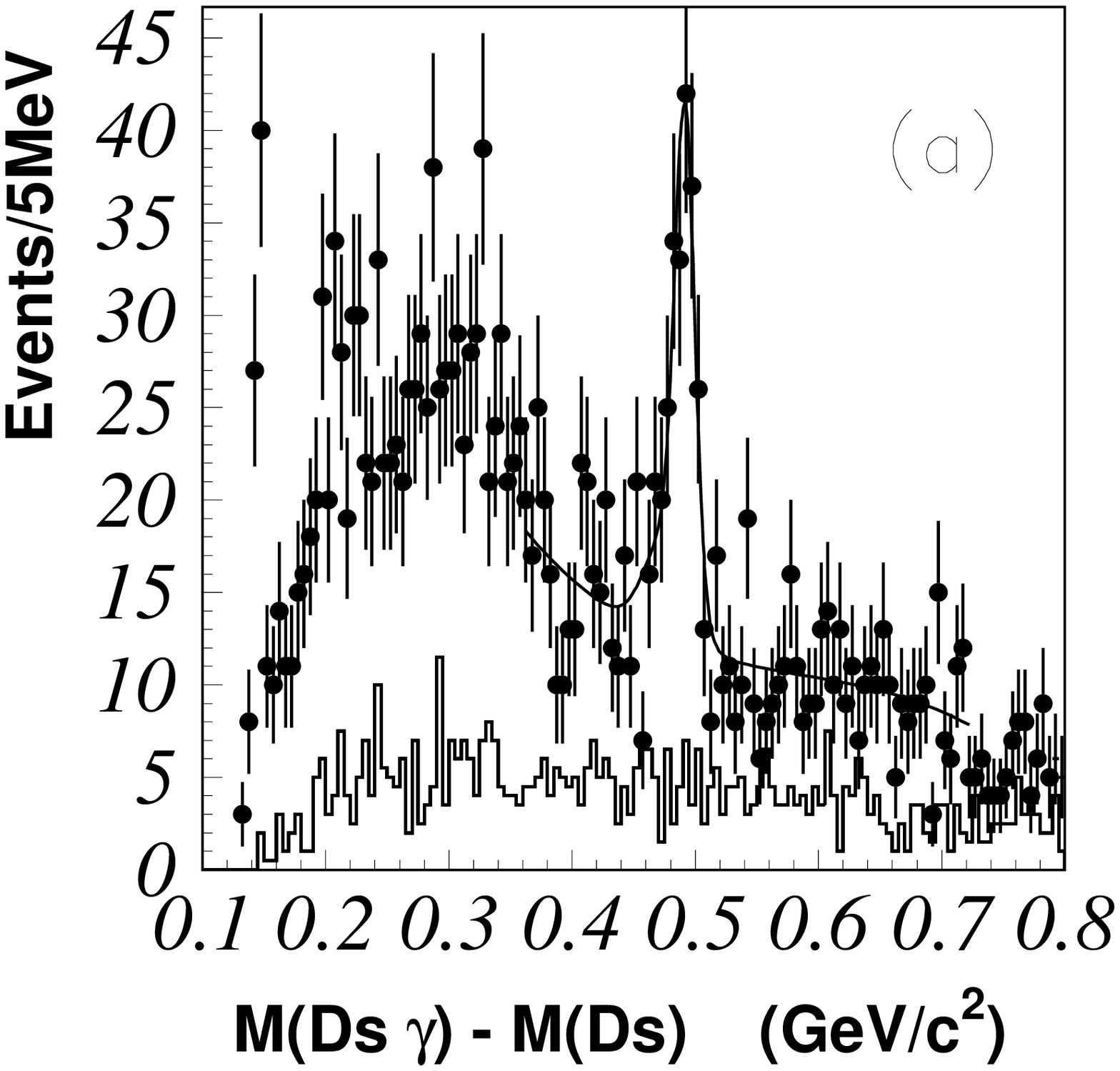}
\end{center}
\end{minipage}
\begin{minipage}{4.25cm}
\begin{center}
\vspace*{-5mm}
\includegraphics[width=4.7cm,clip]{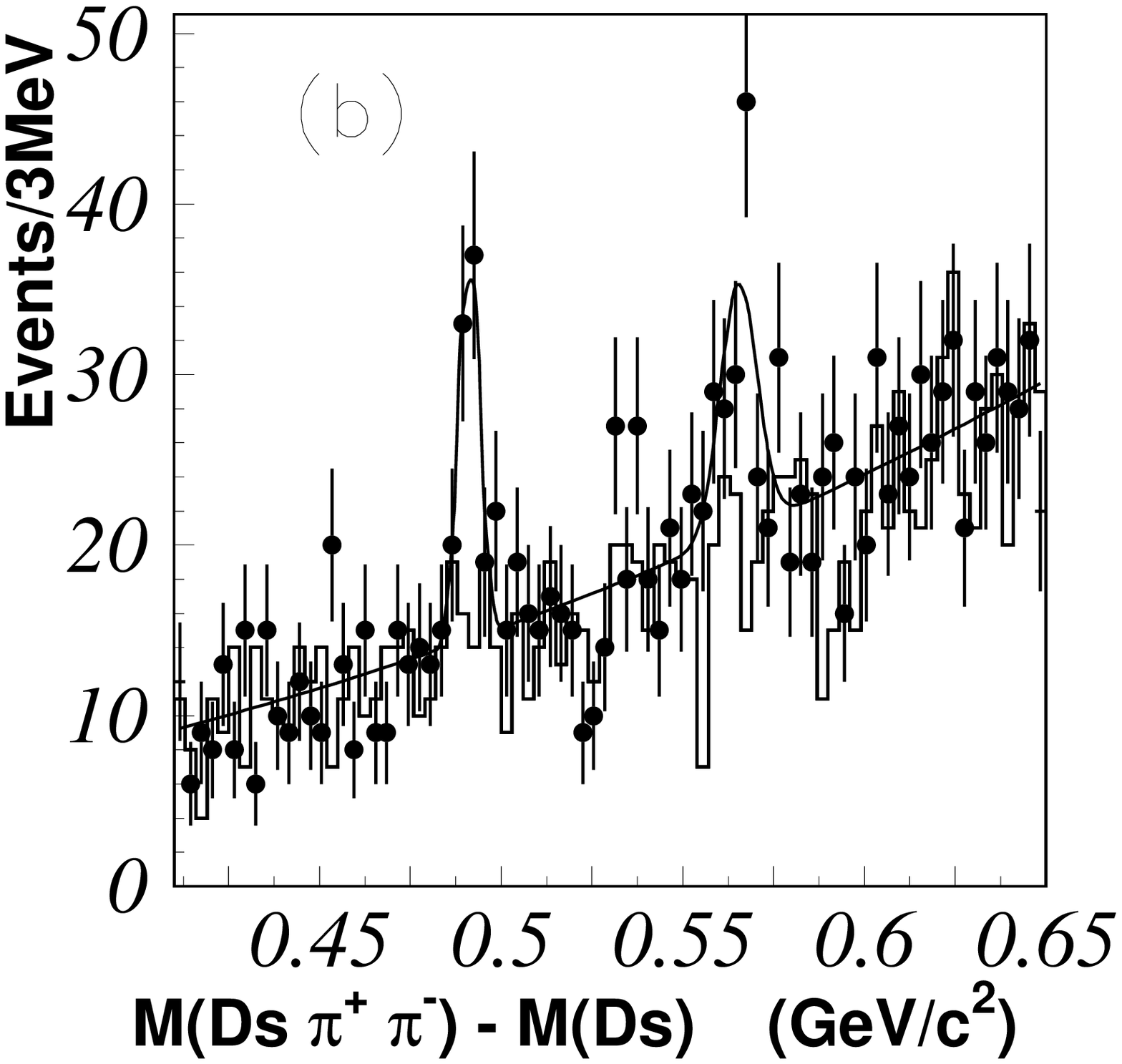}
\hfill
\end{center}
\end{minipage}
 \caption{
(a) The $\Delta M(D_s^+ \gamma)$ distribution. 
The curve is a fit using a double Gaussian for the signal and 
a third-order polynomial for the background. 
(b) The $\Delta M(D_s^+ \pi^+ \pi^-)$ distribution. The
curve is a fit using Gaussian for the signals and a third-order
polynomial for the background.}
\label{radpipi}
\end{figure}

Figure~\ref{radpipi}(a) shows the $\Delta
M(D_s^+ \gamma) = M(D_s^+ \gamma) -M_{D_s^+}$ distribution.  Here
photons are required to have energies greater
than 600~MeV in the CM and those that form a
$\pi^0$ when combined with another photon in the event are not used. 
A clear peak near $\Delta M(D_s^+ \gamma) \sim 490\,\mathrm{MeV}/c^2$,
corresponding to the $D_{sJ}(2457)$, is observed.  
No peak is found in the $D_{sJ}(2317)$ region. 
The $D_s^+$ sideband distribution, shown as a histogram, shows no structure. 
We fit the distribution with a double Gaussian for the signal, which is 
determined from the MC, and a third-order polynomial for the
background. 
The fit yields $152 \pm 18$ (stat) events 
and a 
$\Delta M$ peak at $491.0 \pm 1.3(\mathrm{stat}) \pm 1.9(\mathrm{syst})
\,\mathrm{MeV}/c^2$ 
(corresponding to $M = 2459.5 \pm 1.3(\mathrm{stat}) \pm 2.0(\mathrm{syst})
\,\mathrm{MeV}/c^2$).
The $D_{sJ}(2457)$ mass determined here is consistent with the value 
determined from $D_s^* \pi^0$ decays. 

Using the detection efficiency of 10.2\% for the $D_s^+ \gamma$ decay mode, 
we determine the branching fraction ratio 
$$
\frac{\mathcal{B}(D_{sJ}^+(2457) \to D_s^+ \gamma)}
     {\mathcal{B}(D_{sJ}^+(2457) \to D_s^{*+} \pi^0)}
  = 0.55 \pm 0.13(\mathrm{stat}) \pm 0.08(\mathrm{syst}).
$$
This result, which has a statistical significance of 10$\sigma$, 
is consistent with the first measurement by Belle~\cite{krokovny} 
$0.38 \pm 0.11(\mathrm{stat}) \pm 0.04(\mathrm{syst})$
with $B \to \bar{D} D_{sJ}(2457)$ decays, and  with 
the theoretical predictions~\cite{bardeen},\cite{godfrey}.  
The existence of the $D_{sJ}(2457) \to D_s \gamma$ mode
rules out the $0^{\pm}$ quantum number assignments for the  
$D_{sJ}(2457)$ state.  
For the $D_{sJ}(2317)$, we obtain the upper limit
$$
\frac{\mathcal{B}(D_{sJ}^+(2317) \to D_s^+ \gamma)}
     {\mathcal{B}(D_{sJ}^+(2317) \to D_s \pi^0)}
  \leq 0.05~(\mathrm{90\%~C.L.}).
$$
From the $M(D_s^{*+} \gamma) = M(D_s^{*+} \gamma) -M_{D_s^{*+}}$ 
distribution, we determine the upper limits
$$
\frac{\mathcal{B}(D_{sJ}^+(2317) \to D_s^{*+} \gamma)}
     {\mathcal{B}(D_{sJ}^+(2317) \to D_s \pi^0)}
  \leq 0.18~(\mathrm{90\%~C.L.})~~\mathrm{and}
$$
$$
\frac{\mathcal{B}(D_{sJ}^+(2457) \to D_s^{*+} \gamma)}
     {\mathcal{B}(D_{sJ}^+(2457) \to D_s^{*+} \pi^0)}
  \leq 0.31~(\mathrm{90\%~C.L.}).
$$
Figure~\ref{radpipi}(b) shows the 
$\Delta M(D_s^+ \pi^+ \pi^-) = M(D_s^+ \pi^+ \pi^-) -M_{D_s^+}$ 
distribution.  For pions, 
we require at least one of them to have momentum greater than 
$300\,\mathrm{MeV}/c$ in the CM, one with $P(K/\pi) < 0.1$ and other with 
$P(K/\pi) < 0.9$, and $|M(\pi^+ \pi^-)-M_{K_S}|\geq 15\,\mathrm{MeV}/c^2$. 
A clear peak near $\Delta M(D_s^+ \pi^+ \pi^-) \sim 490\,\mathrm{MeV}/c^2$,
corresponding to the $D_{sJ}(2457)$, is observed.  Evidence of an
additional peak near  $\Delta M(D_s^+ \pi^+ \pi^-) \sim 570\,\mathrm{MeV}/c^2$ 
corresponding to $D_{s1}(2536)$ is also visible. 
No peak is found in the $D_{sJ}(2317)$ region. 
The $D_s^+$ sideband distribution, shown as a histogram, shows no structure. 
We fit the distribution with Gaussians for the signals, which are 
determined from the MC, and a third-order polynomial for the
background. 
The fit yields $59.7 \pm 11.5(\mathrm{stat})$ events and a $\Delta M$ peak 
at $491.4 \pm 0.9(\mathrm{stat}) \pm 1.5(\mathrm{syst})\,
\mathrm{MeV}/c^2$ (corresponding to $M = 2459.9 \pm 0.9(\mathrm{stat}) 
\pm 1.6(\mathrm{syst})\,\mathrm{MeV}/c^2$) for $D_{sJ}(2457)$, 
and  $56.5 \pm 13.4 (\mathrm{stat})$ events for $D_{s1}(2536)$. 
The statistical significance is 5.7$\sigma$ for $D_{sJ}(2457)$, 
and 4.5$\sigma$ for $D_{s1}(2536)$.
This is the first observation of the $D_{sJ}(2457) \to D_s^+ \pi^+ \pi^-$  
decay mode.

The existence of the $D_{sJ}(2457) \to D_s \pi^+ \pi^-$ mode also rules
out the  $0^+$ assignment for $D_{sJ}(2457)$.
Using the detection efficiency of 15.8\% for the $D_s \pi^+ \pi^-$ decay 
mode, we determine the branching fraction ratio 
$$
\frac{\mathcal{B}(D_{sJ}^+(2457) \to D_s^+ \pi^+ \pi^-)}
     {\mathcal{B}(D_{sJ}^+(2457) \to D_s^{*+} \pi^0)}
  = 0.14 \pm 0.04(\mathrm{stat}) \pm 0.02(\mathrm{syst}),
$$
where the systematic error is dominated by the systematic uncertainty
of the $D_{sJ}(2457) \to D_s^{*+} \pi^0$ yield.
We establish the upper limit
$$
\frac{\mathcal{B}(D_{sJ}^+(2317) \to D_s^+ \pi^+ \pi^-)}
     {\mathcal{B}(D_{sJ}^+(2317) \to D_s^+ \pi^0)}
  \leq 4 \times 10^{-3}~(\mathrm{90\%~C.L.}).
$$
Using the detection efficiency of 14.3\% for the 
$D_{s1}(2536) \to D_s \pi^+ \pi^-$ decay mode which assumes the same 
fragmentation function for
the $D_{s1}(2536)$ and $D_{sJ}(2457)$,
we establish the cross section times branching fraction ratio 
\begin{eqnarray}
&&\frac{\sigma(D_{s1}(2536)) \cdot \mathcal{B}(D_{s1}^+(2536) 
\to D_s^+ \pi^+ \pi^-)}
     {\sigma(D_{sJ}(2457)) \cdot \mathcal{B}(D_{sJ}^+(2457) 
\to D_s^+ \pi^+ \pi^-)}
  \nonumber \\ 
 &=&  1.05 \pm 0.32(\mathrm{stat}) \pm 0.06(\mathrm{syst})  \nonumber. 
\end{eqnarray}

In summary, 
we observe radiative and dipion decays of the $D_{sJ}(2457)$ and set upper 
limits on the corresponding decays of the $D_{sJ}(2317)$. 
We determine the $D_{sJ}(2317)$ and $D_{sJ}(2457)$ masses from their decays 
to $D_s^+ \pi^0$ and $D_s^{*+} \pi^0$, respectively, 
and set an upper limit on the decay of $D_{sJ}(2457)$ to $D_s^{+} \pi^0$. 
These results are consistent with the spin-parity assignments for 
the $D_{sJ}(2317)$ and $D_{sJ}(2457)$ of $0^+$ and $1^+$, respectively.

We wish to thank the KEKB accelerator group for the excellent
operation of the KEKB accelerator.
We acknowledge support from the Ministry of Education,
Culture, Sports, Science, and Technology of Japan
and the Japan Society for the Promotion of Science;
the Australian Research Council
and the Australian Department of Education, Science and Training;
the National Science Foundation of China under contract No.~10175071;
the Department of Science and Technology of India;
the BK21 program of the Ministry of Education of Korea
and the CHEP SRC program of the Korea Science and Engineering Foundation;
the Polish State Committee for Scientific Research
under contract No.~2P03B 01324;
the Ministry of Science and Technology of the Russian Federation;
the Ministry of Education, Science and Sport of the Republic of Slovenia;
the National Science Council and the Ministry of Education of Taiwan;
and the U.S.\ Department of Energy.


\begin{thebibliography}{99}
\bibitem{pwave-doublets} 
  {In the heavy $c$-quark approximation, one expects two doublets of 
   $c\bar{s}$ states with quantum numbers $J^P=0^+$, $1^+$ ($j = 1/2$) 
   and $J^P=1^+$, $2^+$ ($j = 3/2$).
   The second doublet has already been observed in $DK$ and 
   $D^*K$ decays~\cite{PDG}.}
\end{thebibliography}
\end{document}